\documentclass[a4paper,10pt]{article}
\usepackage[utf8x]{inputenc}
\usepackage{amsthm,amssymb,amsmath}
\usepackage{graphicx}
\usepackage{epsfig}
\usepackage{slashed}

\newcommand{\oh}{\Omega h^2}
\newcommand{\gev}{\mathrm{GeV}}
\newcommand{\tev}{\mathrm{TeV}}
\newcommand{\sv}{\langle\sigma v\rangle}
\newcommand{\ma}{M_{A^0}}
\newcommand{\mh}{M_{H^0}}
\newcommand{\mhc}{M_{H^\pm}}
\newcommand{\dmn}{M_{N_i}-M_{H^0}}
\newcommand{\mn}[1]{M_{N_{#1}}}

\title{Scalar dark matter and fermion coannihilations in the radiative seesaw model}
\author{Michael Klasen\footnote{michael.klasen@uni-muenster.de}  ~and 
Carlos E. Yaguna\footnote{carlos.yaguna@uni-muenster.de} \\ 
\it \small Institut f\"ur Theoretische Physik, Universit\"at M\"unster,\\
\it \small Wilhelm-Klemm-Stra\ss e 9, D-48149 M\"unster, Germany\\[4mm]
Jos\'e D. Ruiz-\'Alvarez\footnote{Now at IPNL, Lyon, France. jose@gfif.udea.edu.co}, Diego Restrepo\footnote{restrepo@udea.edu.co} ~and Oscar Zapata\footnote{ozapata@fisica.udea.edu.co}\\
\it \small  Instituto de F\'{i}sica, Universidad de Antioquia,\\
\it \small  A.A. 1226, Medell\'{i}n, Colombia}
\date{}

\begin{document}

\maketitle
\vspace*{-10cm}
\begin{flushright}
\texttt{MS-TP-13-03}
\end{flushright}
\vspace*{9cm}
\begin{abstract}
By extending the Standard Model with three right-handed neutrinos ($N_i$) and a second Higgs doublet ($H_2$),  odd under a $Z_2$ symmetry, it is possible to  explain non-zero neutrino masses  and to account for the dark matter. We consider the case where the dark matter is a scalar and  study its coannihilations with the right-handed neutrinos. These coannihilations tend to increase, rather than reduce, the dark matter  density and they modify in a significant way the viable parameter space of the model. In particular,  they allow to satisfy the relic density constraint for dark matter masses well below $500~\gev$. The dependence of the relic density on the relevant parameters of the model, such as the dark matter mass, the mass splitting, and the number of coannihilating fermions, is analyzed in detail. We also investigate, via  a scan over the parameter space, the new viable regions that are obtained when coannihilations are taken into account. Notably, they  feature large indirect detection rates, with $\sv$ reaching values of order $10^{-24}\text{cm}^3\text{s}^{-1}$. Finally, we emphasize that  coannihilation effects analogous to those discussed here can be used to reconcile a thermal freeze-out with a large $\sv$ also in other models of dark matter.
\end{abstract}

\section{Motivation}
Dark matter and neutrino masses provide the only two clear evidences we currently have of physics beyond the Standard Model (SM). Cosmological observations indicate that  about $25\%$ of the energy density of the Universe consists of a new form of matter usually called \emph{dark matter} \cite{Komatsu:2010fb}.  Since none of the known elementary particles satisfy the conditions required to explain the dark matter, new physics should be responsible for it. Even if we still have no clue of what this new physics is,  circumstantial evidence suggests that it lies at the $\tev$ scale \cite{Bertone:2010at}.
The observation of neutrino oscillations \cite{Fukuda:1998mi,Ahmad:2002jz,Araki:2004mb,Adamson:2008zt} implies that neutrinos have non-zero masses --a fact that cannot be accounted for within the SM. In extensions of the SM, neutrino masses can be generated in several ways. Some of them require new physics at very large scales whereas others do not. Normally, dark matter and neutrino masses are treated as separate problems, without any relation to each other. It may be, though, that they are intimately linked and that both originate from physics at the electroweak scale.

A simple model that explicitly realizes this idea was proposed several years ago in \cite{Ma:2006km}. In this model, known as the radiative seesaw model, the SM is extended with a second Higgs doublet and three right-handed neutrinos. A viable dark matter candidate is obtained by assuming that the new fields are odd under a $Z_2$ discrete symmetry, and neutrino masses are generated by one-loop processes mediated by the odd fields, in particular, by  the dark matter candidate.   The dark matter phenomenology of this model has been extensively studied in previous works, see e.g. \cite{Kubo:2006yx,Sierra:2008wj,Gelmini:2009xd,Suematsu:2009ww,Suematsu:2011va,Schmidt:2012yg,Hu:2012az,Kashiwase:2013uy,Kashiwase:2012xd}.  None of these, however, has considered the  possible effect of coannihilations with the right-handed neutrinos for scalar dark matter. That is precisely what we do in this paper. We will see that these coannihilations tend to increase the relic density of dark matter and may modify in a significant way the viable parameter space of the model. Specifically, they allow to reduce the lower bound on the mass of the dark matter particle from $500~\gev$ down to about $100~\gev$.   We study in detail how these coannihilation processes affect the calculation of the relic density for different values of the parameters of the model, and we analyze the resulting viable parameter space. Remarkably, we find that, due to coannihilations, the indirect detection rate of dark matter is strongly enhanced within the new viable regions.

The rest of the paper is organized as follows. The next section introduces the radiative seesaw model and fixes our notation. In section \ref{sec:coan} we discuss the coannihilation formalism and how it can be applied  to  our model. Our first results are presented in section \ref{sec:res}, where we analyze  the dependence of the coannihilation effects on the dark matter density for different sets of parameters. In section \ref{sec:imp} we investigate the new viable parameter space that is obtained when coannihilations are taken into account. In particular, we demonstrate that it features much larger values of $\sv$. Section \ref{sec:dis} brings up some possible extensions of our work and section \ref{sec:con} presents our conclusions.

\section{The radiative seesaw model}
The radiative seesaw model \cite{Ma:2006km} is a minimal extension of the Standard Model that simultaneously accounts for neutrino masses and for the dark matter. It includes three right-handed neutrinos $N_i$ ($i=1,2,3$) and one additional Higgs doublet $H_2$. These new fields are assumed to be odd under a new (and exact) $Z_2$ discrete symmetry under which all  the SM fields are even. The new terms that appear in the Lagrangian of this model are
\begin{equation}
 \mathcal{L}_N=\bar N_ii\slashed \partial P_R N_i+\left(D_\mu H_2\right)^\dagger\left(D^\mu H_2\right)-\frac{M_i}{2}\bar N_i^cP_RN_i+h_{\alpha i}\bar \ell_\alpha H_2^\dagger P_RN_i+\text{h.c.},
\end{equation}
and 

\begin{align}
 -\mathcal{L}_{H_1,H_2}& =  \mu_{1}^2 H_1^\dagger H_1+ \mu_{2}^2 H_2^\dagger H_2 +\frac{\lambda_1}{2} \left(H_1^\dagger H_1\right)^2 +\frac{\lambda_2}{2} \left(H_2^\dagger H_2\right)^2\\
&+\lambda_3 \left(H_1^\dagger H_1\right)\left(H_2^\dagger H_2 \right)+\lambda_4\left(H_1^\dagger H_2\right)\left(H_2^\dagger H_1\right)+\frac{\lambda_5}{2}\left(H_1^\dagger H_2\right)^2
+\text{h.c.},\nonumber
\end{align}
where $H_1$ is the SM Higgs doublet. Had  only $H_2$ (but not $N_i$) been added we would have obtained the so-called inert doublet model \cite{Barbieri:2006dq,LopezHonorez:2006gr}, a minimal extension of the Standard Model that can explain the dark matter. Thus, in certain regions of the parameter space, the radiative seesaw can  be seen as an extension of the inert doublet model that accounts also for neutrino masses. It is also important to notice that $N_i$, being odd under $Z_2$, are not the same right-handed neutrinos that appear in the usual seesaw mechanism. They do not give rise, for instance, to a Dirac mass term for the neutrinos.   

The new scalar sector of this model contains four physical states: two charged states, $H^{\pm}$, and two neutral ones, $H^0$ and $A^0$. Their masses are given by
\begin{align}
 \mhc^2&= \mu_2^2+\lambda_3v^2 ,\nonumber \\
\mh^2&= \mu_2^2+(\lambda_3+\lambda_4+\lambda_5)v^2 ,\nonumber \\
\ma^2&= \mu_2^2+(\lambda_3+\lambda_4-\lambda_5)v^2\,, 
\end{align}
where $v$ is the vacuum expectation value of $H_1$.  We take as the free parameters of the scalar sector the physical masses ($\mh,\ma,\mhc$) and the coupling $\lambda\equiv \lambda_3+\lambda_4+\lambda_5$.

Majorana neutrino masses in this model are generated at one-loop by the exchange of $H^0,A^0$ and $N_i$ \cite{Ma:2006km}. The resulting mass matrix is given by
\begin{equation}
 \left(m_\nu\right)_{\alpha\beta}\simeq \sum_{i=1}^3\frac{2 \lambda_5 h_{\alpha i}h_{\beta i} v^2}{(4\pi)^2 M_i} I\left(\frac{M_i^2}{M_0^2}\right),
\end{equation}
where $M_i$ are the masses of the right-handed neutrinos, $M_0^2\simeq \mu_{2}^2+(\lambda_3+\lambda_4) v^2$, and the loop function is given by
\begin{equation}
 I(x)=\frac{x}{1-x}\left(1+\frac{x\log x}{1-x}\right).
\end{equation}

Compatibility with present neutrino data can be achieved if the matrix of Yukawa couplings $h_{\alpha i}$ has the following structure \cite{Suematsu:2009ww}
\begin{equation}
 h_{\alpha i}=\begin{pmatrix}
               0 & 0 & h_3'\\ h_1 & h_2 & h_3 \\ h_1 & h_2 & -h_3
              \end{pmatrix}.
\label{eq:hs}
\end{equation}
From it, it follows that $\theta_{23}=\pi/4$, $\theta_{13}=0$ and $\tan\theta_{12}=\frac{1}{\sqrt 2}h_3'/h_3$. Current data, see e.g. \cite{Tortola:2012te}, then implies $h_3'/h_3\approx 0.95$.  The relation between neutrino masses and the parameters $h_i$ is as follows
\begin{equation}
 (h_1^2+h_2^2)\Lambda_1\simeq \frac{\sqrt{\Delta m^2_\text{atm}}}{2},\quad h_3^2\Lambda_3\simeq \frac{\sqrt{\Delta m^2_\text{sol}}}{3}
\label{eq:h2}
\end{equation}
where
\begin{equation}
 \Lambda_i=\frac{2 \lambda_5 v^2}{(4\pi)^2 M_i} I\left(\frac{M_i^2}{M_0^2}\right)
\label{eq:lambda}
\end{equation}
and we have taken $h_i$ to be real. Notice, therefore, that only one parameter, from those appearing in equation (\ref{eq:hs}), remains undetermined by the experimental data. We take that free parameter to be $h_2$. $h_1$ and $h_3$ are then calculated, for given masses of the odd particles, from equations (\ref{eq:h2}) and (\ref{eq:lambda}). In any case, as we will see later, our results do not depend on the specific choice made in equation (\ref{eq:hs}).

In this model, one-loop diagrams mediated by the odd particles give rise to lepton flavor violating processes, such as $\mu\to e\gamma$ and $\tau\to \mu\gamma$, which  may impose important restrictions on  the parameter space of the model. It turns out, though, that in the region of interest to us, the Yukawa couplings are small and the experimental bounds on lepton flavor violation are easily satisfied.

Regarding dark matter, in this model the lightest odd particle can either be a fermion (one of the three right-handed neutrinos) or a scalar ($H^0$ or $A^0$). The former case has been studied in~\cite{Kubo:2006yx,Sierra:2008wj,Gelmini:2009xd,Suematsu:2009ww,Suematsu:2011va,Schmidt:2012yg,Hu:2012az} and the latter in \cite{Kashiwase:2013uy,Kashiwase:2012xd}. In this paper, we focus on the case where the dark matter candidate is a scalar, $H^0$. Naively, one might expect the resulting dark matter phenomenology to be similar to that of the inert doublet model, but that is not necessarily true. As we will show, $H^0$-$N_i$ coannihilations may play an important role in the determination of the relic density, modifying the viable regions and the prospects for the detection of dark matter. 

\section{Coannihilations and their effect on the relic density}
\label{sec:coan}
The observed dark matter density is believed to be the result of a freeze-out in the early Universe. This freeze-out process can be modified by the presence of additional particles that are close in mass to the dark matter particle --what is usually known as \emph{coannihilations} \cite{Griest:1990kh}. Coannihilations may lead both to an increase or a decrease of the final abundance of the lightest stable particle, the dark matter candidate. In the context of supersymmetry, for example, coannihilations usually cause a suppression of the relic abundance.  Since the neutralino is bino-like in most of these models,  it  annihilates rather inefficiently in the early Universe and for that reason $\Omega_\chi$ is typically too high. It turns out though that, in certain regions of the parameter space, the neutralino  becomes almost degenerate with a stau  so that stau-neutralino coannihilations become relevant.  Because staus annihilate more efficiently than bino-like neutralinos, the effective coannihilation cross section is larger than the neutralino annihilation cross section and the final relic density is lower than without stau coannihilations, making it easier to satisfy the dark matter bound. 

The opposite behavior is observed in scenarios with Universal Extra Dimensions (UED) \cite{Hooper:2007qk}. In UED, the spectrum of Kaluza-Klein states is highly degenerate so coannihilations are expected to play an important role. It has been shown that, for realistic spectra, the ultimate effect of these coannihilations is to increase the relic abundance of the lightest Kaluza-Klein particle, reducing the dark matter mass required to satisfy the relic density constraint \cite{Servant:2002aq,Kong:2005hn,Burnell:2005hm}.  Within supersymmetry, it is also possible to find examples where the effect of coannihilations is to enhance rather than reduce  the relic density.   In \cite{Edsjo:2003us}, it was shown that if bino-like neutralinos annihilate through the resonant exchange of a heavy Higgs, slepton coannihilations give rise to an increase in the relic density.  In \cite{Profumo:2006bx},  models where the neutralino is higgsino-like or wino-like were considered and it was shown that slepton coannihilations not only lead to an increase in the relic density but also to an enhancement in the predicted indirect detection signals.

Let us now briefly review, following \cite{Griest:1990kh,Profumo:2006bx}, the coannihilation formalism and use it to obtain an estimate of the expected effect within the radiative seesaw model. The effective annihilation cross section for a system of $n$ coannihilating particles is given by
\begin{equation}
 \sigma_\text{eff}=\sum_{i,j=1}^n\sigma_{ij}\frac{g_ig_j}{g^2_\text{eff}} (1+\Delta_i)^{3/2}(1+\Delta_j)^{3/2} e^{-x(\Delta_i+\Delta_j)},
\end{equation}
where $\Delta_i=(m_i-m_\chi)/m_\chi$ is the relative mass splitting between the $i$-th coannihilating particle and the lightest stable particle $\chi$, $x=m_\chi/T$, $\sigma_{ij}$ denotes the annihilation cross section of particles $i$ and $j$ into Standard Model particles, $g_i$ stands for the number of internal degrees of freedom
of particle $i$ and
\begin{equation}
 g_\text{eff}=\sum_{i=1}^ng_i(1+\Delta_i)^{3/2}e^{-x\Delta_i}.
\end{equation}
Here, we are interested in the case where the dark matter particle, $H^0$, has an effective pair annihilation cross section much larger than that of its coannihilating partners, the right-handed neutrinos $N_i$. This situation  arises \emph{naturally} within the radiative seesaw model. In fact, $H^0$ annihilates efficiently into $W^+W^-$ and $Z^0Z^0$ via gauge interactions whereas the interactions of the right-handed neutrinos are determined by the  Yukawa couplings and can easily be suppressed. It must be kept in mind, however, that the $H^0$ effective annihilation cross section without $N_i$ coannihilations is actually the result of coannihilations within the scalar sector. As we will see, the mass splitting of $A^0$ and $H^\pm$ with $H^0$ is always small within the viable regions, so those coannihilations are unavoidable. If we denote by $\sigma_{H^0H^0}=\sigma_\text{eff}^0$ the effective $H^0$ annihilation cross section without right-handed neutrino coannihilations then we have that $\sigma_{H^0H^0}\gg \sigma_{H^0N_i}$, $\sigma_{N_iN_i}$, where $\sigma_{H^0N_i}$, $\sigma_{N_iN_i}$ respectively indicate the $N_i$ coannihilation and self-annihilation effective cross sections.  Denoting with $g_\text{eff}^0$ the effective degrees of freedom when the $N_i$ are much heavier than $H^0$, the new effective total annihilation cross section $\sigma_\text{eff}^{N_i}$ can be written as a function of the effective number of degrees of freedom $g_{eff}^{N_i}$ including the $N_i$ as
\begin{equation}
 \sigma_\text{eff}^{N_i}\sim \sigma_\text{eff}\left(\frac{g_\text{eff}^0(x_\text{f.o.})}{g_\text{eff}^{N_i}(x_\text{f.o.})}\right)^2
\end{equation}
where $x_\text{f.o.}$ corresponds to temperatures around the $H^0$ freeze-out. Since $g_{eff}^{N_i}(x_\text{f.o.})$ is larger than $g_\text{eff}^0(x_\text{f.o.})$, the effective annihilation cross section decreases with the inclusion of $N_i$ coannihilations --and the dark matter density increases accordingly. In the limit where $\mh-\mn{i}$ becomes negligible ($\Delta_i\to 0$) one can write
\begin{equation}
 \frac{\Omega^{N_i}}{\Omega^0}\approx \left(\frac{g_0+g_{N_i}}{g_0}\right)^2
\end{equation}
for the ratio between the relic density including $N_i$ coannihilations ($\Omega^{N_i}$) and that without doing so ($\Omega^0$). This approximate expression depends only on the internal degrees of freedom that take part in the coannihilation process. In our case, there are four scalar degrees of freedom ($H^0$, $A^0$, $H^\pm$) so $g_0=4$. Each right-handed neutrino has two degrees of freedom and we could have coannihilations with $1$, $2$ or $3$ of them, so $g_{N_i}$ can respectively take the values $2$, $4$ or $6$. The ratio $\Omega^{N_i}/\Omega^0$ is then equal to $9/4$ for 1 coannihilating neutrino, $4$ for two coannihilating neutrinos, and $25/4$ for $3$ coannihilating neutrinos. These numbers, being obtained for a negligible mass splitting between $N_i$ and $H^0$, actually represent an upper bound to this ratio, so we expect to find smaller effects in   more realistic situations. From this simple discussion we can already conclude, in any case,  that coannihilations with right-handed neutrinos can increase the $H^0$ relic density by up to a factor $6$ or so.

In the next section we evaluate numerically the effect of coannihilations on the relic density. To that end, we have implemented the radiative seesaw model into micrOMEGAs \cite{Belanger:2010gh}, which automatically takes into account all possible coannihilation effects.

\section{Results for the relic density}
\label{sec:res}

\begin{figure}[t]
\begin{center} 
\includegraphics[scale=0.37]{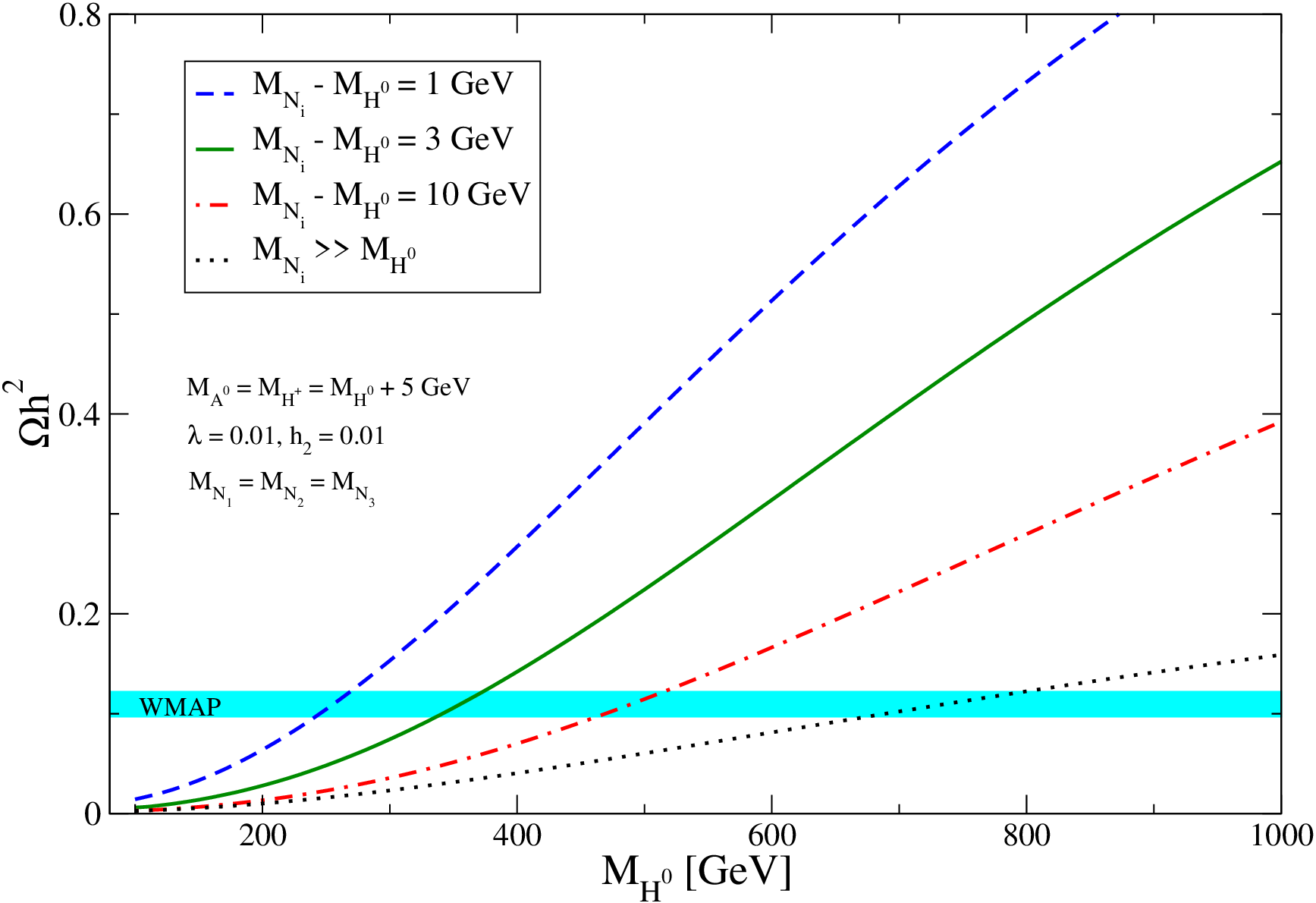}
\caption{The relic density as a function of $\mh$ for different values of $\dmn$. In this figure we have set $\lambda=0.01$, $h_2=0.01$, $\ma=\mhc=\mh+5~\gev$ and we have assumed that the three fermions have the same mass: $\mn{1}=\mn{2}=\mn{3}$. Notice that $\oh$ decreases with increasing $\dmn$.\label{fig:omegadmn}}
\end{center}
\end{figure}

In this section, the effect of $N_i$ coannihilations on the relic density of $H^0$ is numerically studied. Specifically, we analyze how $\oh$ depends on the $N_i$-$H^0$ mass splitting, on the value of the neutrino Yukawa couplings ($h_2$), and on the number of coannihilating neutrinos.  We also determine some  viable regions and study their variation with the parameters of the model. 

When coannihilations with $N_i$ are negligible, the dark matter phenomenology of this model reduces to that of the well-known inert doublet model in the heavy mass regime \cite{LopezHonorez:2006gr,Hambye:2009pw}. It is useful, therefore, to keep in mind the main two features of this latter model. First, the mass splitting between the scalars is necessarily very small along the viable regions. Second, the viable mass range starts at $\mh\gtrsim 500~\gev$. For smaller values of $\mh$, $\Omega_{H^0}$ is always below the observed range.  We will demonstrate that these features are modified in the presence of coannihilations with the right-handed neutrinos.

\begin{figure}[t]
\begin{center} 
\includegraphics[scale=0.4]{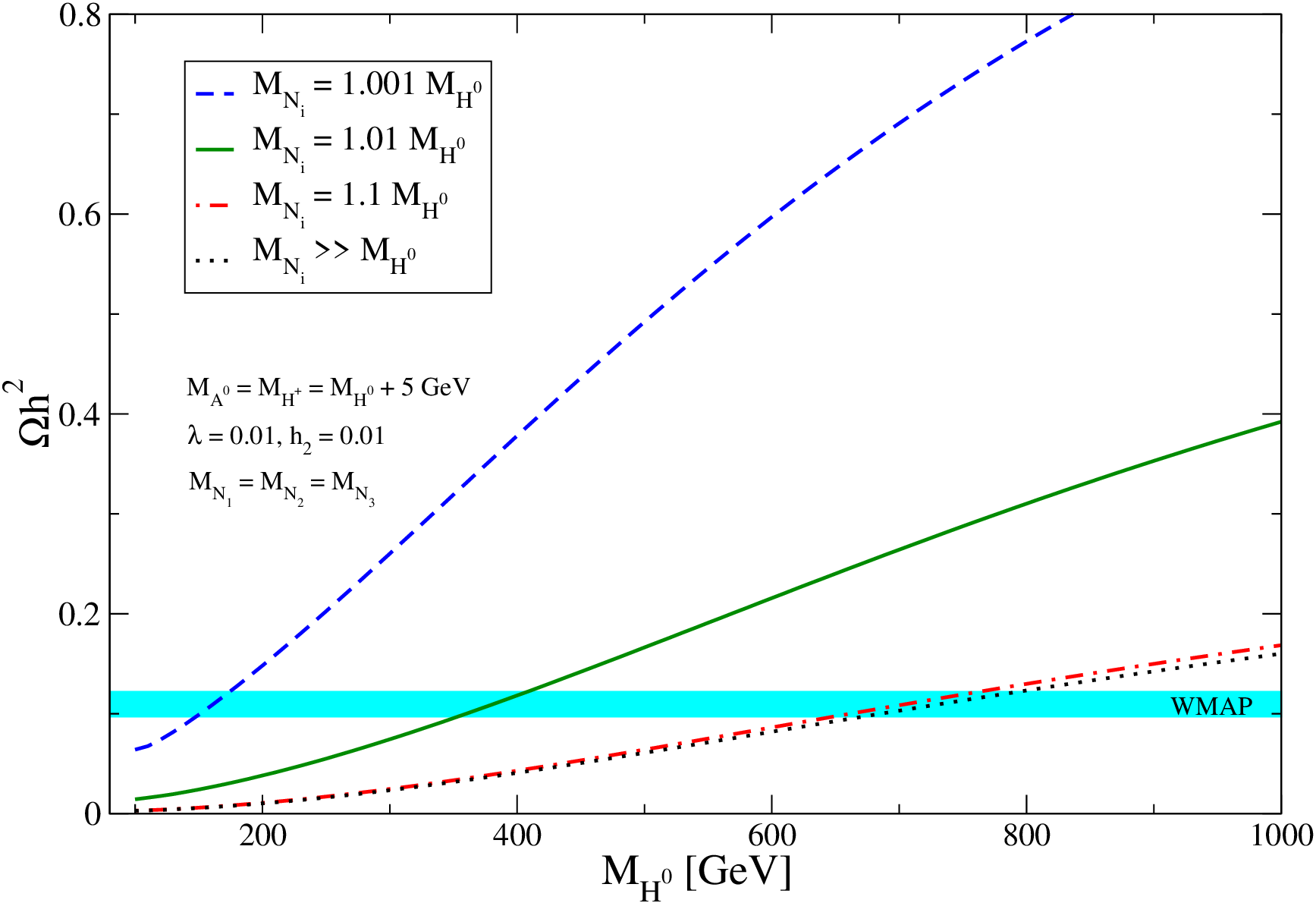}
\caption{The relic density as a function of $\mh$ for different values of $\mn{i}/\mh$. In this figure we have set $\lambda=0.01$, $h_2=0.01$, $\ma=\mhc=\mh+5~\gev$ and we have assumed that the three fermions have the same mass: $\mn{1}=\mn{2}=\mn{3}$. Notice that $\oh$ decreases with increasing $\mn{i}/\mh$.\label{fig:omegadmnf}}
\end{center}
\end{figure}

Figure \ref{fig:omegadmn} shows the relic density as a function of $\mh$ for several values of the mass difference: $\dmn=1~\gev$ (dashed line), $3~\gev$ (solid line), $10~\gev$ (dash-dotted line). For comparison, the case where coannihilation effects are negligible, $\mn{i}\gg\mh$, is also shown (dotted line). The horizontal band shows  the WMAP range for $\oh$ \cite{Komatsu:2010fb}. We see that coannihilation effects increase the relic density and consequently  reduce the value of $\mh$ that is compatible with WMAP data.  As  the mass splitting is reduced, coannihilation effects become more relevant and the increase in the relic density is larger.  The viable value of $\mh$, for instance,  goes from about $750~\gev$ for  $\mn{i}\gg\mh$ (no coannihilations) to about $250~\gev$ for $\dmn=1~\gev$.   

\begin{figure}[tb]
\begin{center} 
\includegraphics[scale=0.4]{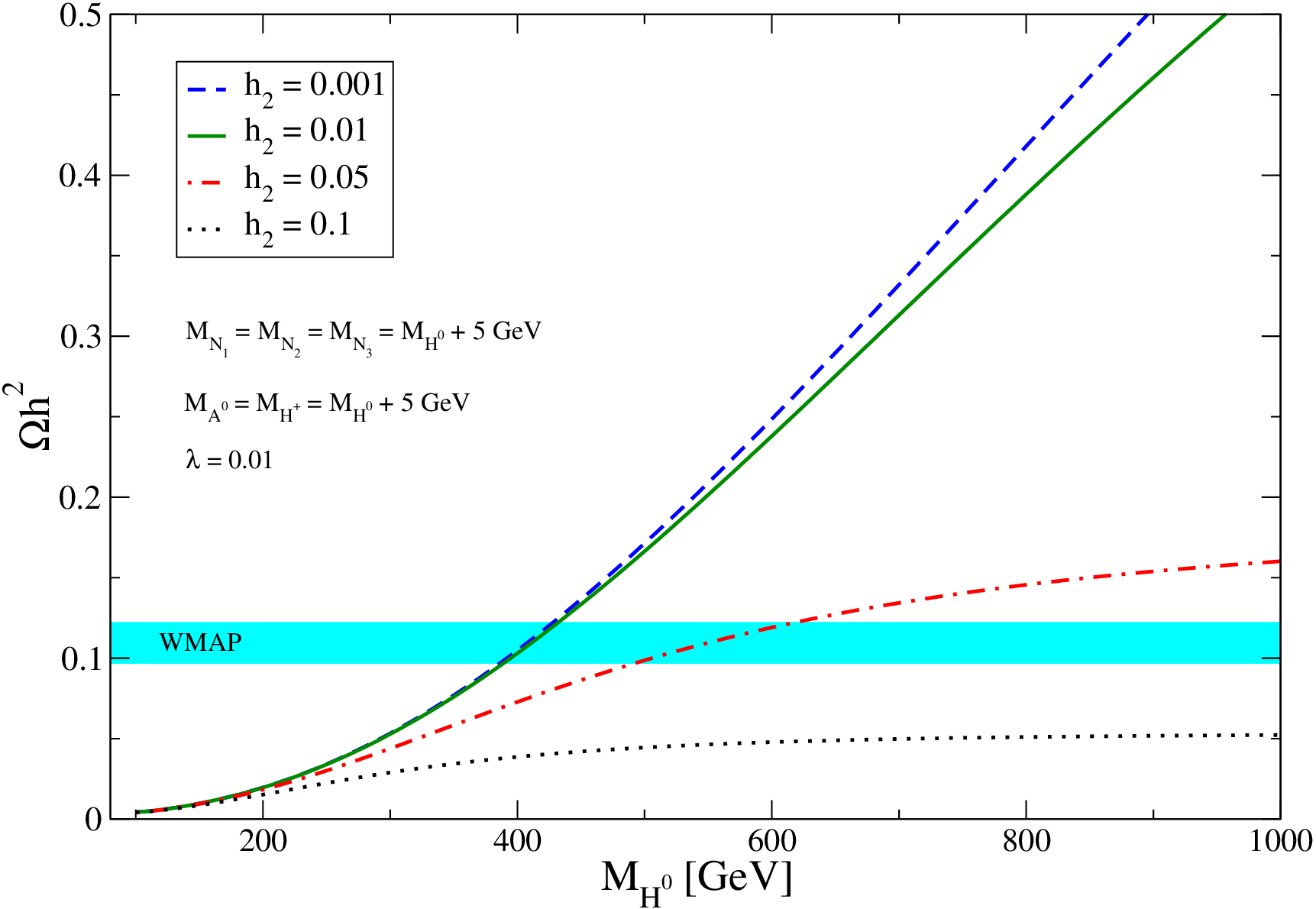}
\caption{The relic density as a function of $\mh$ for different values of $h_2$. In this figure we have set $\lambda=0.01$, $\ma=\mhc=\mh+5~\gev$ and we have assumed that the three fermions have the same mass: $\mn{1}=\mn{2}=\mn{3}=\mh+5~\gev$. \label{fig:omegayuk}}
\end{center}
\end{figure}

To better understand the degree of degeneracy between $N_i$ and $\mh$ that is required to obtain a certain effect on the relic density, it is useful to look at the dependence of $\oh$ with the relative mass difference, $(\dmn)/\mh$ --see figure \ref{fig:omegadmnf}. Notice that a $10\%$ degeneracy (dash-dotted line) produces almost no visible effect, and that in that case $\mh$ is compatible with the observations for   $\mh\sim 750~\gev$ --the same value obtained without coannihilations (dotted line). A $1\%$ degeneracy (solid line) reduces the viable value of  $\mh$ to about $400~\gev$, and a $0.1\%$ degeneracy lowers it further to about $150~\gev$.   

\begin{figure}[tb]
\begin{center} 
\includegraphics[scale=0.4]{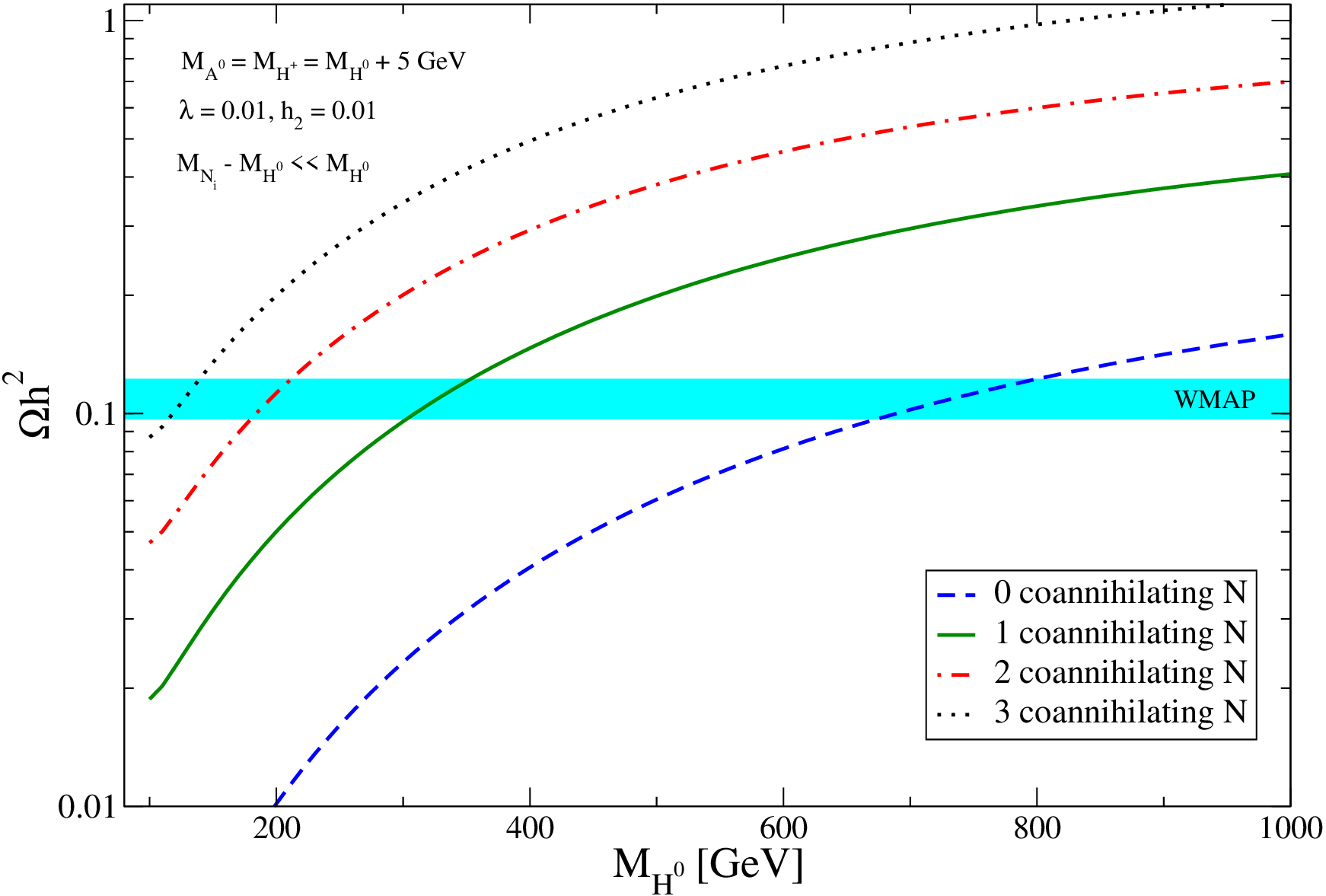}
\caption{The relic density as a function of $\mh$ for different numbers of coannihilating $N$. In this figure we have set $\lambda=0.01$, $h_2=0.01$, $\ma=\mhc=\mh+5~\gev$, and we have assumed that the coannihilating particles have a negligible mass splitting: $\mn{i}-\mh\ll\mh$. \label{fig:omegandeg}}
\end{center}
\end{figure}

The neutrino Yukawa couplings also affect the relic density, as they determine the strength of the $H^0$-$N_i$ cross section. We expect, in particular, that only for small yukawas the right-handed neutrinos will behave as additional degrees of freedom and increase the relic density. Figure \ref{fig:omegayuk} shows $\oh$ as a function of $\mh$ for different values of $h_2$. For $h_2=0.1$ (dotted line) we notice that the relic density is always below the WMAP bound.  For $h_2=0.05$ (dash-dotted line) $\oh$ is slightly larger and it becomes possible to satisfy the dark matter constraint in the range $500<\mh/\gev<600$. For $h_2=0.01$ (solid line) the viable value of $\mh$ is smaller, about  $400~\gev$. That same value is obtained for $h_2=0.001$ (dashed line). In fact, these last two lines only deviate from each other at large values of $\mh$, where $\oh$ is well above the observed range. Considering even smaller values of $h_2$ would have no further effects on the relic density.  

\begin{figure}[tb]
\begin{center} 
\includegraphics[scale=0.4]{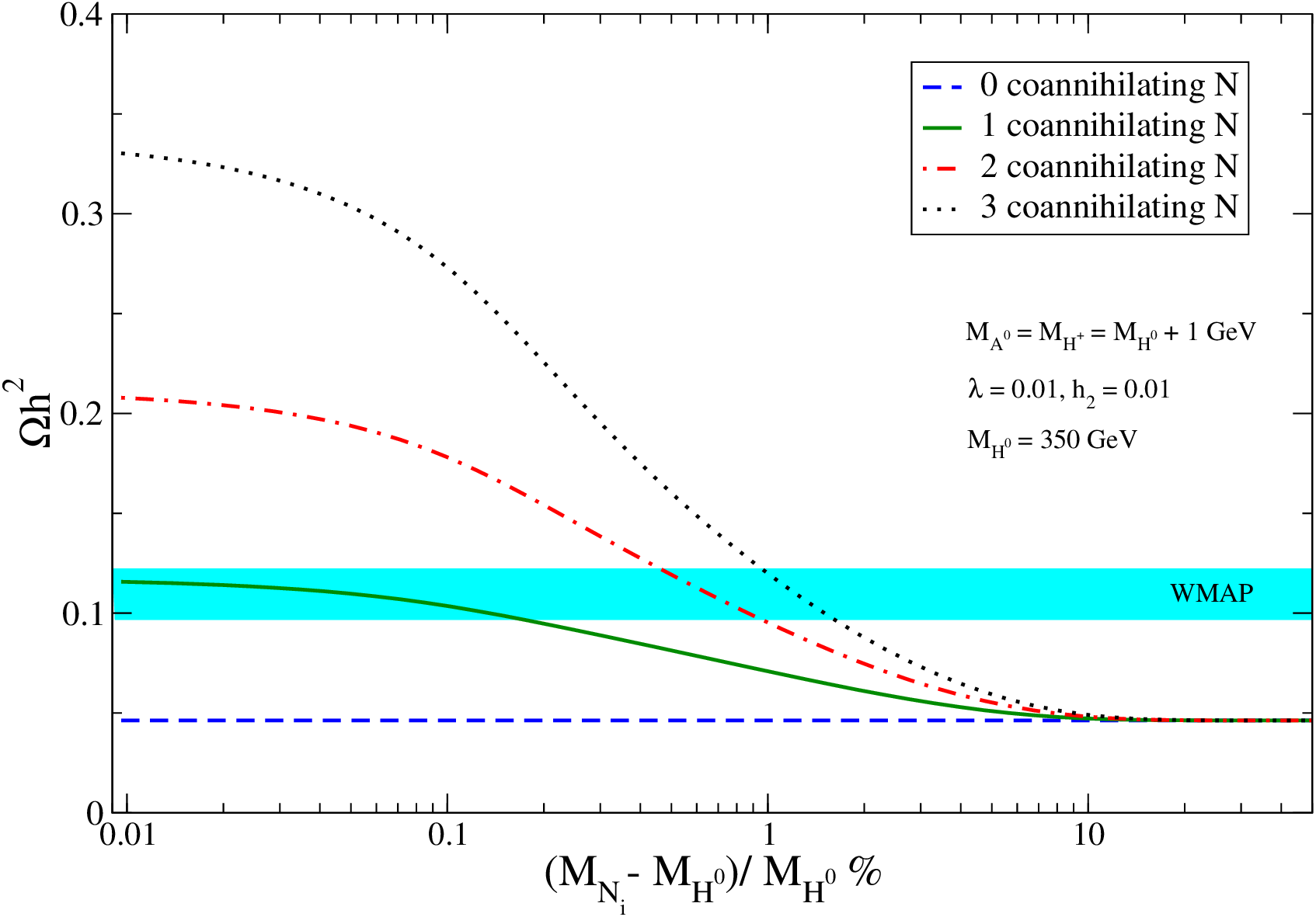}
\caption{The relic density as a function of the ratio $(\dmn)/\mh$ for different numbers of coannihilating $N$. In this figure we have set $\lambda=0.01$, $h_2=0.01$, $\ma=\mhc=\mh+1~\gev$, and $\mh=350~\gev$. \label{fig:omegafixedmh0}}
\end{center}
\end{figure}

In all the previous figures we have assumed that the three right-handed neutrinos have the same mass, $\mn{1}=\mn{2}=\mn{3}$, so they all coannihilate simultaneously with $H^0$. It may well be, though, that only $1$ or $2$ of them actually take part in the coannihilation process. In the next figures, we study how the predicted relic density depends on the number of coannihilating $N$. Figure \ref{fig:omegandeg} shows the relic density as a function of $\mh$ for $0$, $1$, $2$, and $3$ coannihilating fermions. In this figure we have assumed that the mass splitting between the coannihilating neutrinos and $H^0$ is negligible, $\mn{i}-\mh\ll\mh$, whereas the mass of the non-coannihilating neutrino is much larger than that of $H^0$. First of all, notice that, as expected, the relic density increases with the number of coannihilating particles or rather with the number of coannihilating degrees of freedom. If there are $0$ coannihilating neutrinos (dashed line) we reproduce the known result from the inert doublet model and we observe that $\mh$ should be around $750~\gev$ to account for the dark matter. For $1$ coannihilating neutrino, the viable mass moves down to $\mh\sim 300~\gev$, a value that can never be reached within the inert doublet model. For $2$ and $3$ coannihilating neutrinos we obtain respectively $\mh\sim 200~\gev$ and $\mh\sim 150~\gev$ for the values that are compatible with WMAP. Thus, the number of coannihilating neutrinos plays a very important role in the determination of the $H^0$ relic density.

\begin{figure}[tb]
\begin{center} 
\includegraphics[scale=0.4]{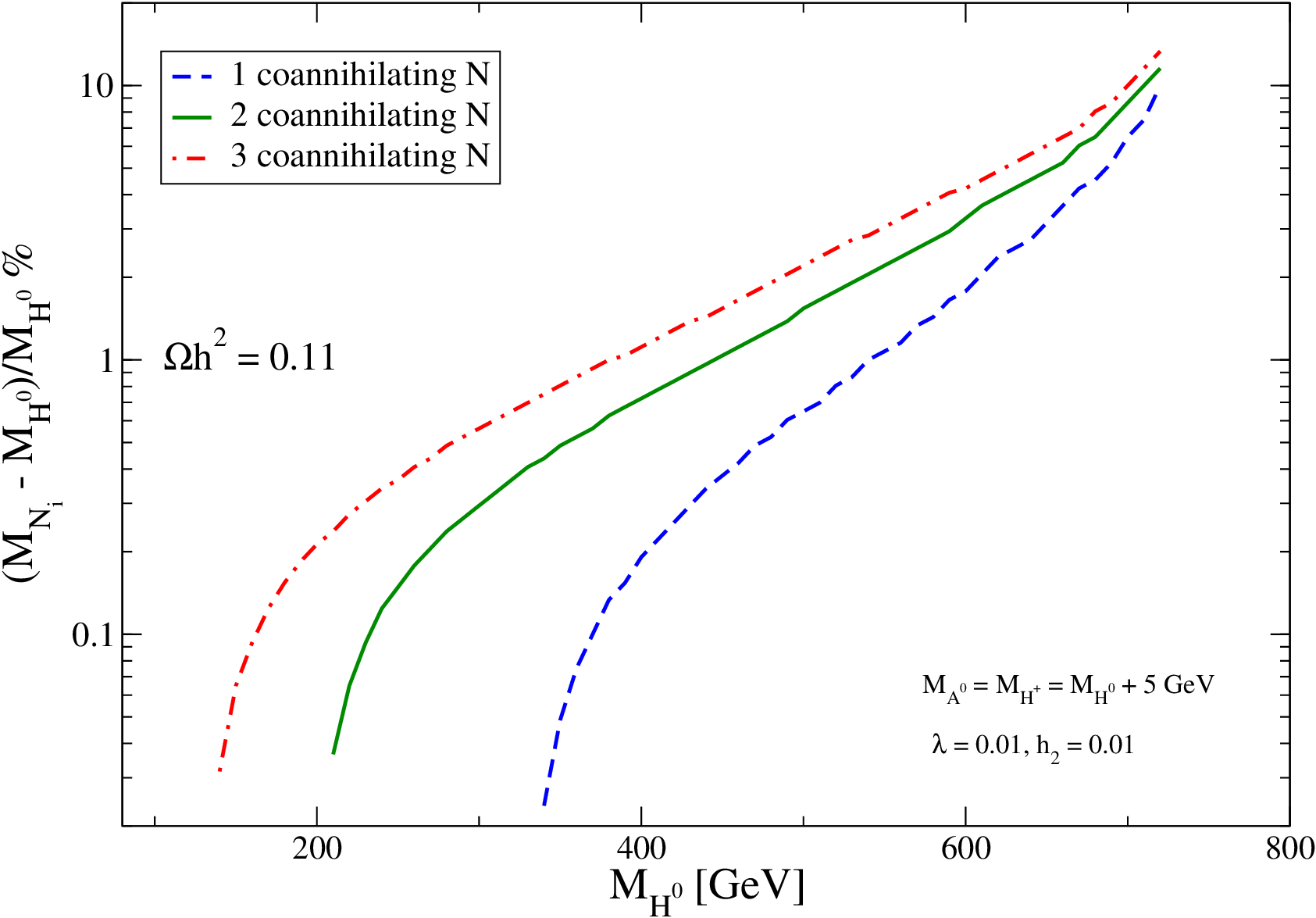}
\caption{The regions in the plane ($\mh$,$(\dmn)/\mh$) that are consistent with the relic density constraint for different numbers of coannihilating $N$. In this figure we have set $\lambda=0.01$, $h_2=0.01$, $\ma=\mhc=\mh+5~\gev$. \label{fig:omegalevel}}
\end{center}
\end{figure}

Figure \ref{fig:omegafixedmh0} shows $\oh$ as a function of the relative mass difference, ($\dmn$)$/\mh$, for all the possible values of the coannihilating neutrinos and a fixed value of the dark matter mass, $\mh=350~\gev$. First of all notice that the coannihilation effect completely disappears when ($\dmn$)$/\mh$ reaches about $10\%$. For higher values, the relic density is not affected by coannihilations so it becomes independent of the number of coannihilating $N$. It is also clear from this figure that $\oh$ cannot be made arbitrarily large by reducing the mass splitting. There is a limit to this trend, for $\oh$ reaches an asymptotic value at very small ($\dmn$)$/\mh$. In that limit, the numerical results coincide with the estimates obtained in section \ref{sec:coan}. For $0$ coannihilating neutrinos (dashed line) we observe from figure \ref{fig:omegafixedmh0} that the dark matter constraint cannot be satisfied --as expected, because that is the inert Higgs model limit and $\mh=350~\gev<500~\gev$. For $1$ coannihilating neutrino, the right relic density is obtained for a mass degeneracy below the $0.1\%$ level. The required mass degeneracy lies between $0.5\%$ and $1\%$ for $2$ coannihilating neutrinos, and between $1\%$ and $2\%$ for $3$ coannihilating neutrinos.

It is also important to look at how the viable regions, those that satisfy the dark matter constraint, depend on the mass splitting and on the number of coannihilating particles. Figure \ref{fig:omegalevel} shows such regions in the plane ($\mh$,$(\dmn)/\mh$) for $1$, $2$ and $3$ coannihilating neutrinos. Notice that these regions start respectively at about $150~\gev$, $200~\gev$, and $350~\gev$, and that in all cases the initial degeneracy is below $0.1\%$. As $\mh$ increases so does the required degree of degeneracy, reaching about $1\%$ respectively for $350~\gev$, $400~\gev$, and $550~\gev$. When  $(\dmn)/\mh$ becomes of order $10\%$ the coannihilation effect fades away and all three lines converge. 

\section{Implications}
\label{sec:imp}
We have already observed that $H^0$-$N_i$ coannihilations can modify  the predicted relic density of $H^0$, giving rise to  new viable regions which may differ considerably from those obtained without coannihilations. In this section we are going to study such viable regions in more detail. We will see that they indeed allow for significant differences in the spectrum and result in modified prospects for the detection of dark matter.

In this analysis, we will make use of a scan over the entire parameter space of this model. We have generated four large sample of models, each with a different number of coannihilating $N$. In each case, we have varied  the parameters of the model  within the following ranges:
\begin{align}
100~\gev &< \mh<1~\tev\\
\mh &< \ma<\mh+40~\gev\\
\mh &< \mhc<\mh+40~\gev\\
\mh &< \mn{i}<\mh+40~\gev\\
10^{-5} &< \lambda<10^{-1}\\
10^{-6} &< h_2<10^{-1}\,.
\end{align}
Here, $\mn{i}$ denotes the masses of the coannihilating $N$. The non-coannihilating neutrinos  were given a mass much larger than $\mh$. From the randomly generated models, we have selected those that satisfy the dark matter constraint to obtain the sample of viable models that we are going to analyze next.

\begin{figure}[tb]
\begin{center} 
\begin{tabular}{cc}
\includegraphics[scale=0.21]{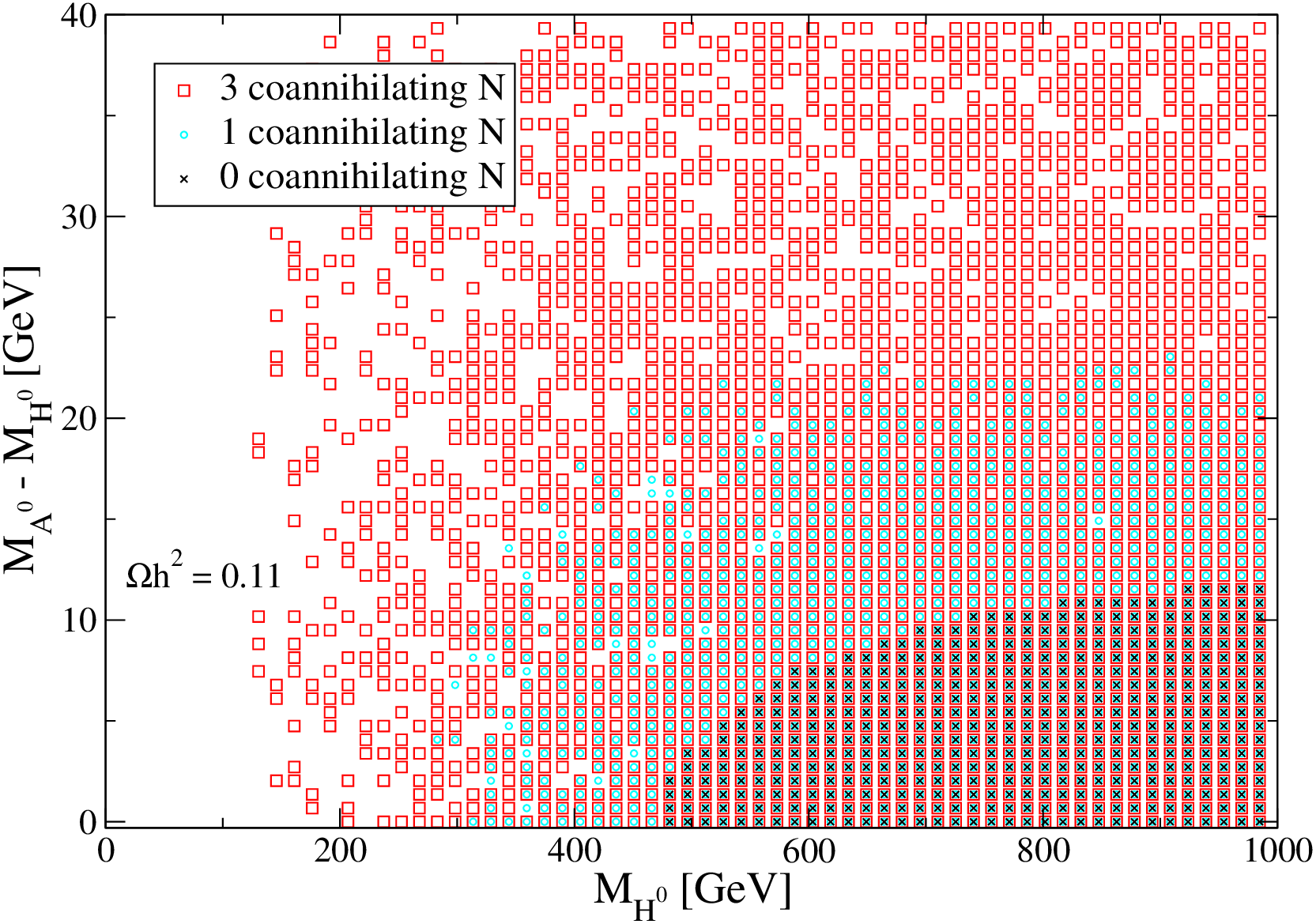} & \includegraphics[scale=0.21]{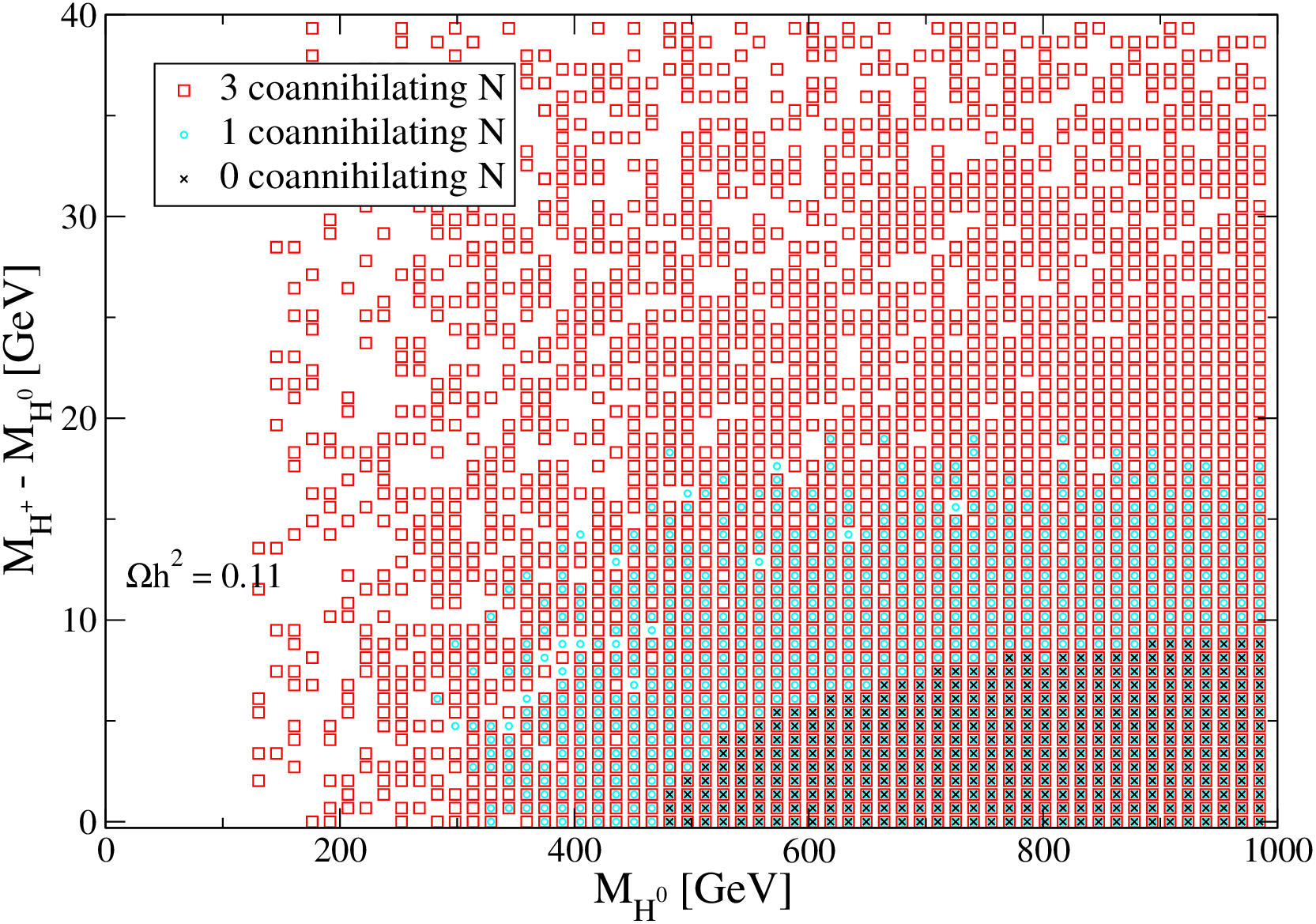}
\end{tabular}
\caption{A scatter plot showing the mass difference between the scalar particles versus $\mh$ for different numbers of coannihilating $N$: $3$ (red), $1$ (blue) and $0$ (black). \label{fig:mh0vsmasses}}
\end{center}
\end{figure}

Figure \ref{fig:mh0vsmasses} shows a scatter plot of the mass difference between the scalar particles versus $\mh$ for different numbers of coannihilating $N$: $3$ (red), $1$ (blue) and $0$ (black). The left panel displays $\ma-\mh$ and the right panel $\mhc-\mh$. When there is no coannihilations (black points), we notice that there are no points below $\mh\sim 500~\gev$ and that the mass splitting is always very small, $\lesssim 10~\gev$. These are the well-known results for the inert doublet model. If there is one coannihilating neutrino (blue circles), the viable mass range starts at about $\mh\sim300~\gev$ and the mass splitting can be larger, reaching $\ma-\mh\sim 25~\gev$ and $\mhc-\mh\sim 20~\gev$. In the case where the three neutrinos coannihilate (red squares), the whole  range of $\mh$ becomes viable and the mass splitting can reach the maximum values we examined, $\sim 40~\gev$.  It is clear, therefore, that  the viable regions when $H^0$-$N_i$ coannihilations are included may feature much smaller dark matter masses (down to $\mh\sim 100~\gev$ or so) and larger values for the mass splitting between the new scalar states.

\begin{figure}[tb]
\begin{center} 
\begin{tabular}{cc}
\includegraphics[scale=0.21]{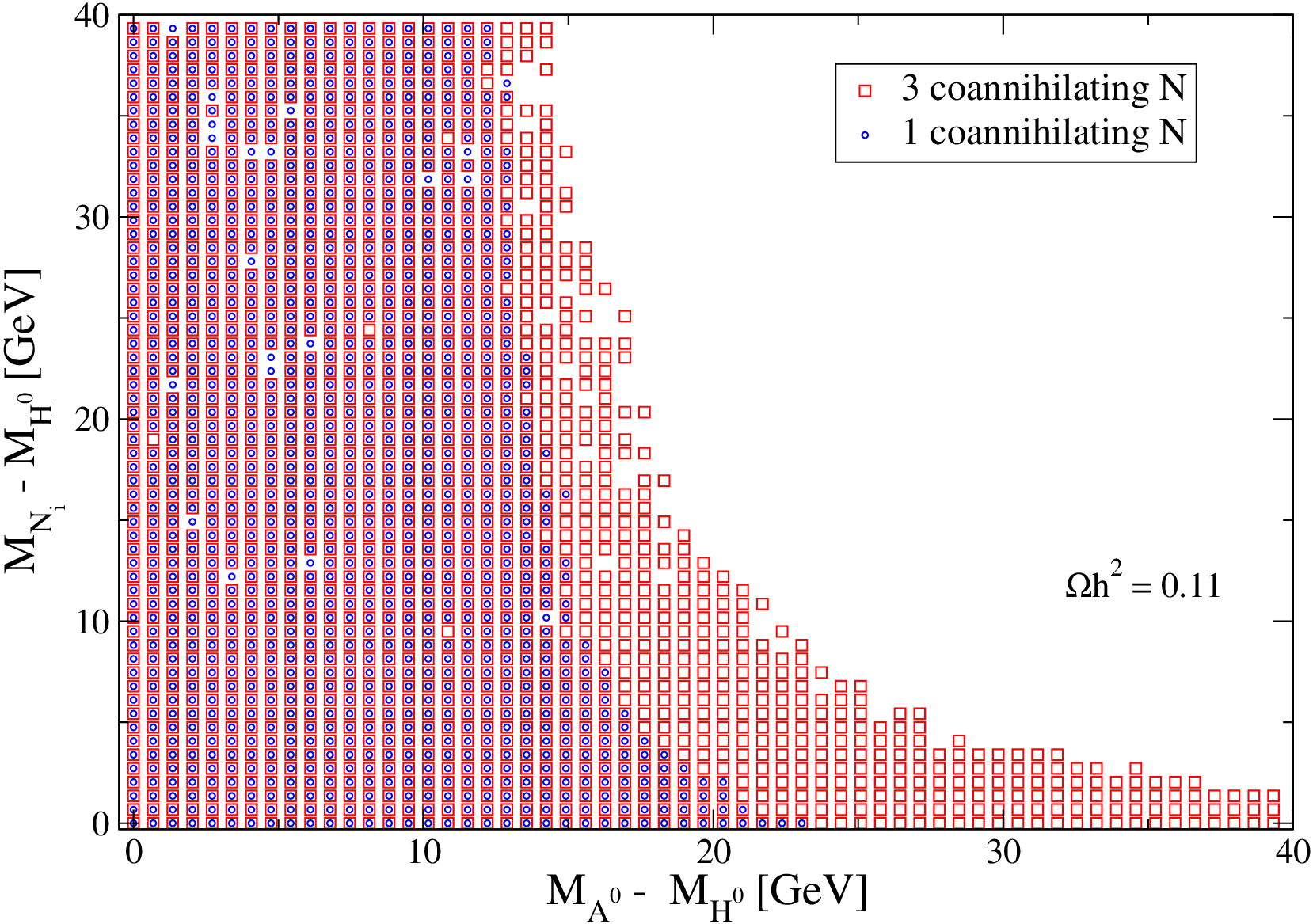} & \includegraphics[scale=0.21]{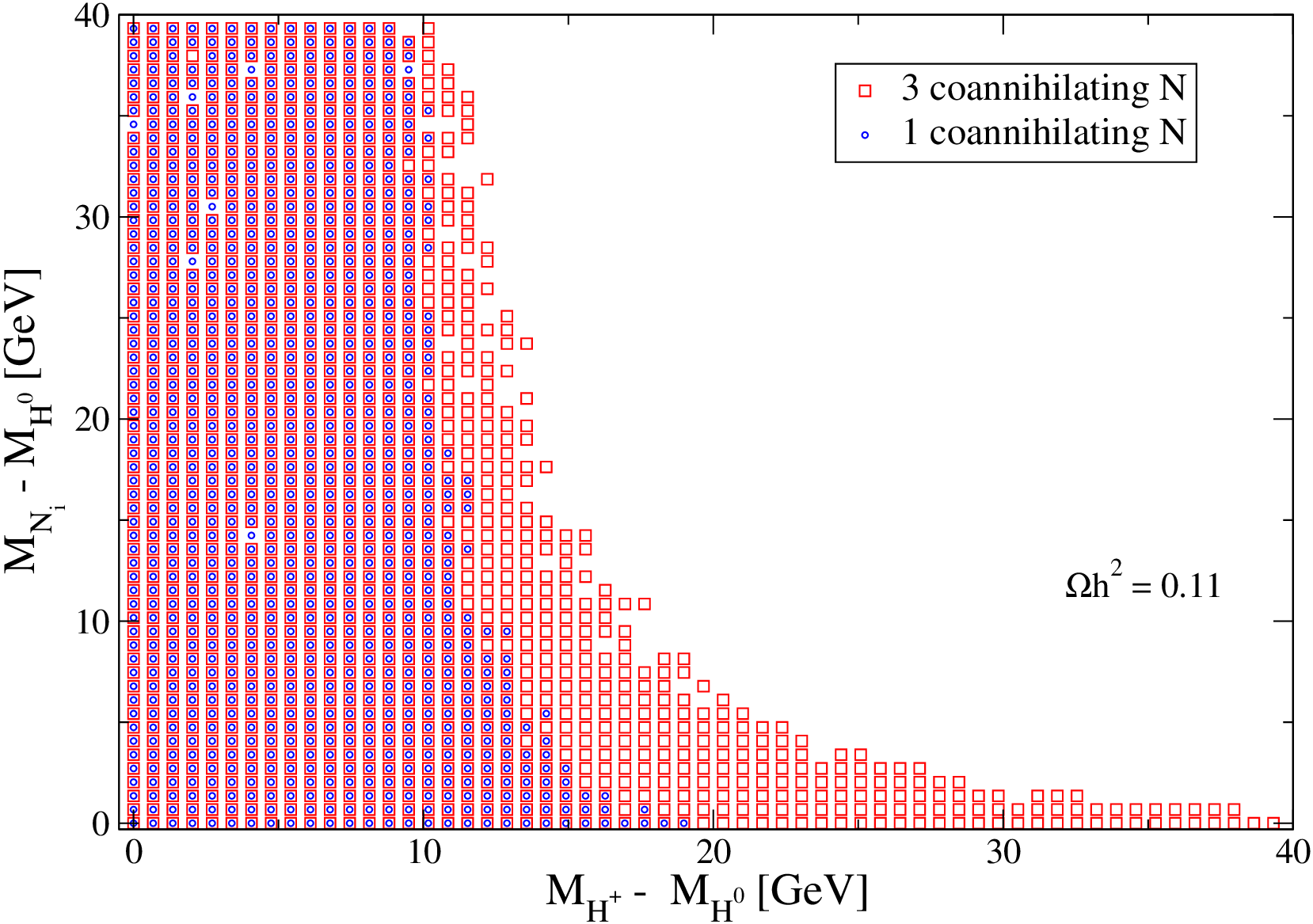}
\end{tabular}
\caption{Scatter plots showing the mass difference between the scalar particles versus $\dmn$ for two different numbers of coannihilating $N$: $3$ (red), $1$ (blue). \label{fig:dmvsdmn}}
\end{center}
\end{figure}

What allows $\ma-\mh$ and $\mhc-\mh$ to be larger are the coannihilations with the right-handed neutrinos, as illustrated by figure \ref{fig:dmvsdmn}. It displays the scalar mass differences versus $\dmn$ for $1$ (blue circles) and $3$ (red squares) coannihilating neutrinos. Notice, from both panels, that $\ma-\mh$ and $\mhc-\mh$ can become relatively large only when $\dmn$ becomes small. Thus, one way or another there must always be some particle highly degenerate with $\mh$.

\begin{figure}[tb]
\begin{center} 
\includegraphics[scale=0.4]{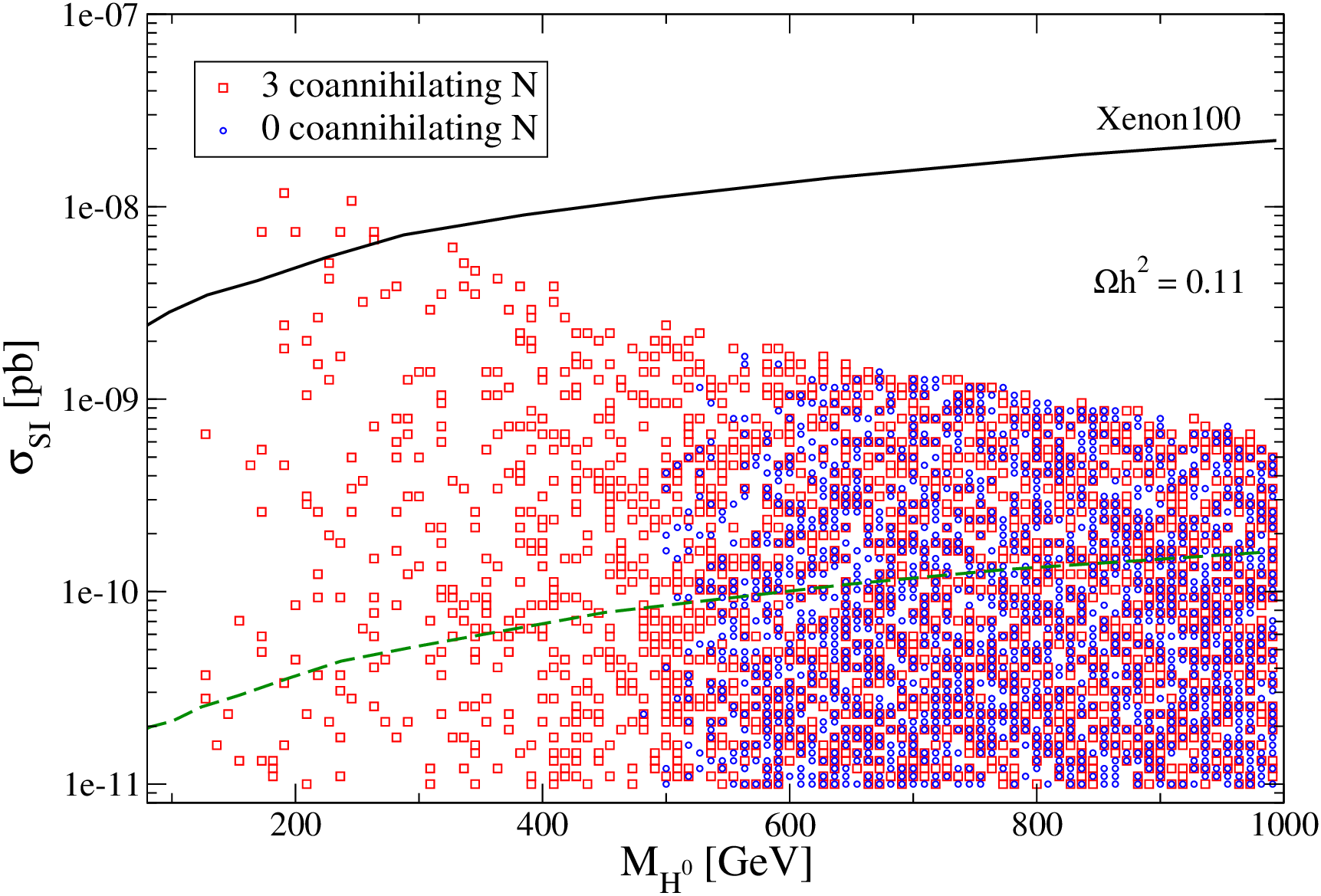} 
\caption{Scatter plot of the spin-independent direct detection cross section ($\sigma_\mathrm{SI}$) versus $\mh$ for $0$ (blue circles) and $3$ (red squares) coannihilating neutrinos. In this figure we show only those models featuring $\sigma_\mathrm{SI}>10^{-11}$ pb at tree-level.  The solid line corresponds to the present bound from XENON100, while the dashed line corresponds to the expected sensitivity of XENON1T. \label{fig:mh0vsdd}}
\end{center}
\end{figure}

Let us now look at how the prospects for the detection of dark matter are modified in the presence of coannihilations. The dark matter spin-independent direct detection cross section ($\sigma_\mathrm{SI}$) in this model is determined at tree-level by a Higgs mediated diagram and is proportional to $\lambda^2$--just as for the inert doublet model. As pointed out recently in \cite{Klasen:2013btp}, however, the one-loop corrections to this cross section can be large and  substantially modify the predicted value of $\sigma_\mathrm{SI}$. In particular, they  always bring its value within the reach of future direct detection experiments, $\sim 10^{-11}$ pb. These results are not affected by the presence of coannihilating right-handed neutrinos. Their main effect is simply to allow for viable models with smaller masses for the  dark matter particle, a region where direct detection experiments have a larger sensitivity. This fact is illustrated in figure \ref{fig:mh0vsdd}, which shows a scatter plot of $\sigma_\mathrm{SI}$ versus $\mh$ for $0$ (blue circles) and $3$ (red squares) coannihilating neutrinos. The solid line shows the present bound from XENON100 \cite{Aprile:2012nq} and the dashed line the expected sensitivity of XENON1T. Notice that whereas for $3$ coannihilating neutrinos a handful of models with masses around $200~\gev$ is already ruled out by the experimental bound, not a single model is currently above the limit for $0$ coannihilating neutrinos--they all lie at least one order of magnitude below it.  In the future this difference will remain. The smallest $\sigma_\mathrm{SI}$ that could be excluded by XENON1T is about $10^{-10}$ pb for $0$ coannihilating neutrinos but $3\times10^{-11}$ pb for $3$ coannihilating neutrinos. Coannihilations, therefore, may have an important impact in the direct detection prospects of dark matter within this model.

\begin{figure}[tb]
\begin{center} 
\includegraphics[scale=0.4]{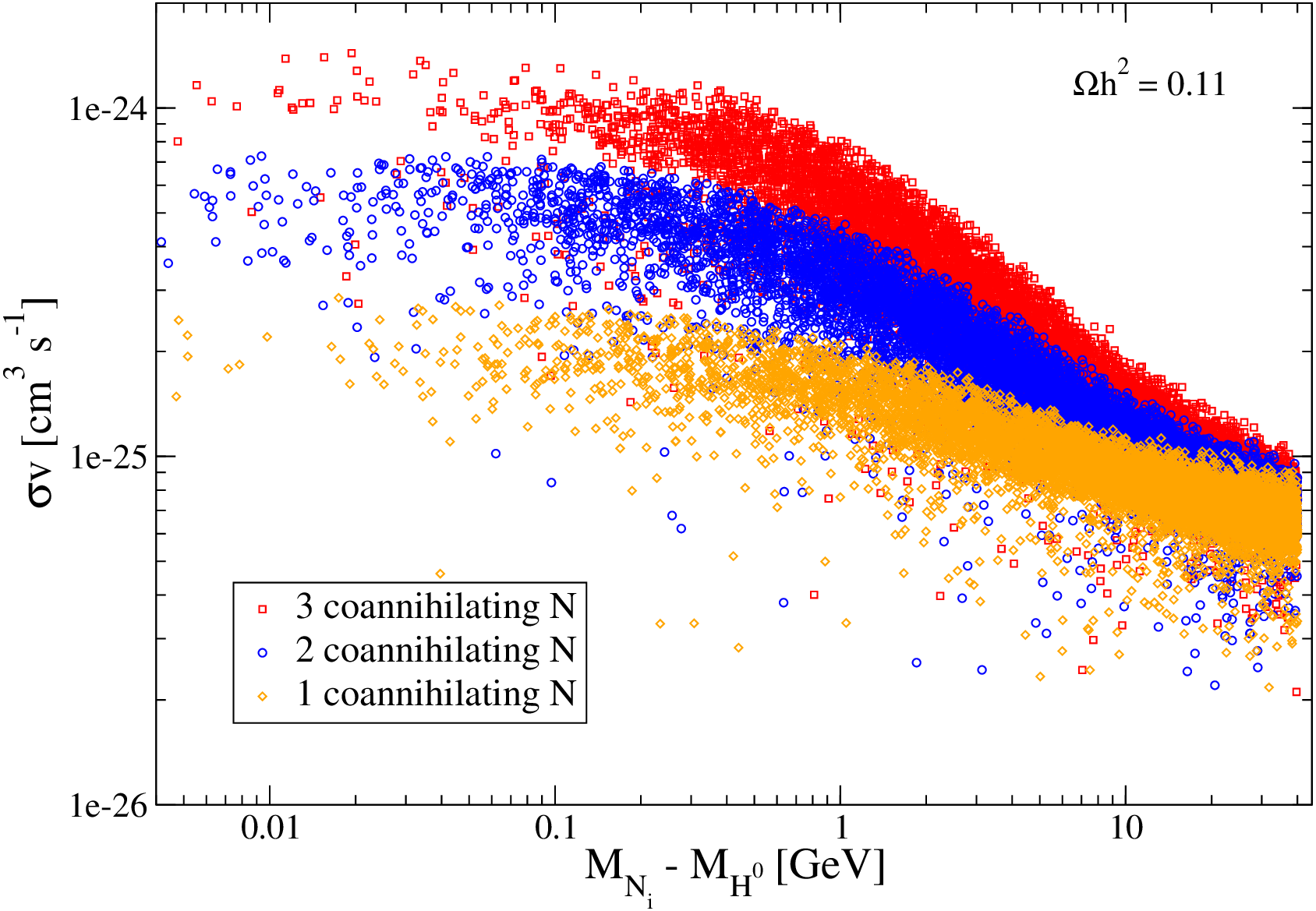} 
\caption{Scatter plots of $\dmn$ versus $\sv$ for different numbers of coannihilating $N$: $1$ (orange diamonds), $2$ (blue circles) and $3$ (red squares). \label{fig:sigmavdmn}}
\end{center}
\end{figure}

What about indirect detection? Well, it turns out that  $N_i$ coannihilations may have even more significant implications for the indirect detection of dark matter,  as they allow to reconcile a thermal freeze-out with large values of $\sv$, and therefore with enhanced indirect detection signals. Figure \ref{fig:sigmavdmn} shows $\sv$ as a function of $\dmn$ for different number of coannihilating neutrinos. We see that the higher the number of coannihilating neutrinos the larger $\sv$ can be. In fact, $\sv$ goes from a maximum of  $2\times 10^{-25}\mathrm{cm^3s^{-1}}$ for 1 degenerate neutrino  to a maximum of $10^{-24}\mathrm{cm^3s^{-1}}$ for $3$ degenerate neutrinos. It is also clear that $\sv$  decreases as the $N_i$-$H^0$ increases, so the largest $\sv$ correspond to the smallest $\dmn$.

\begin{figure}[tb]
\begin{center} 
\includegraphics[scale=0.4]{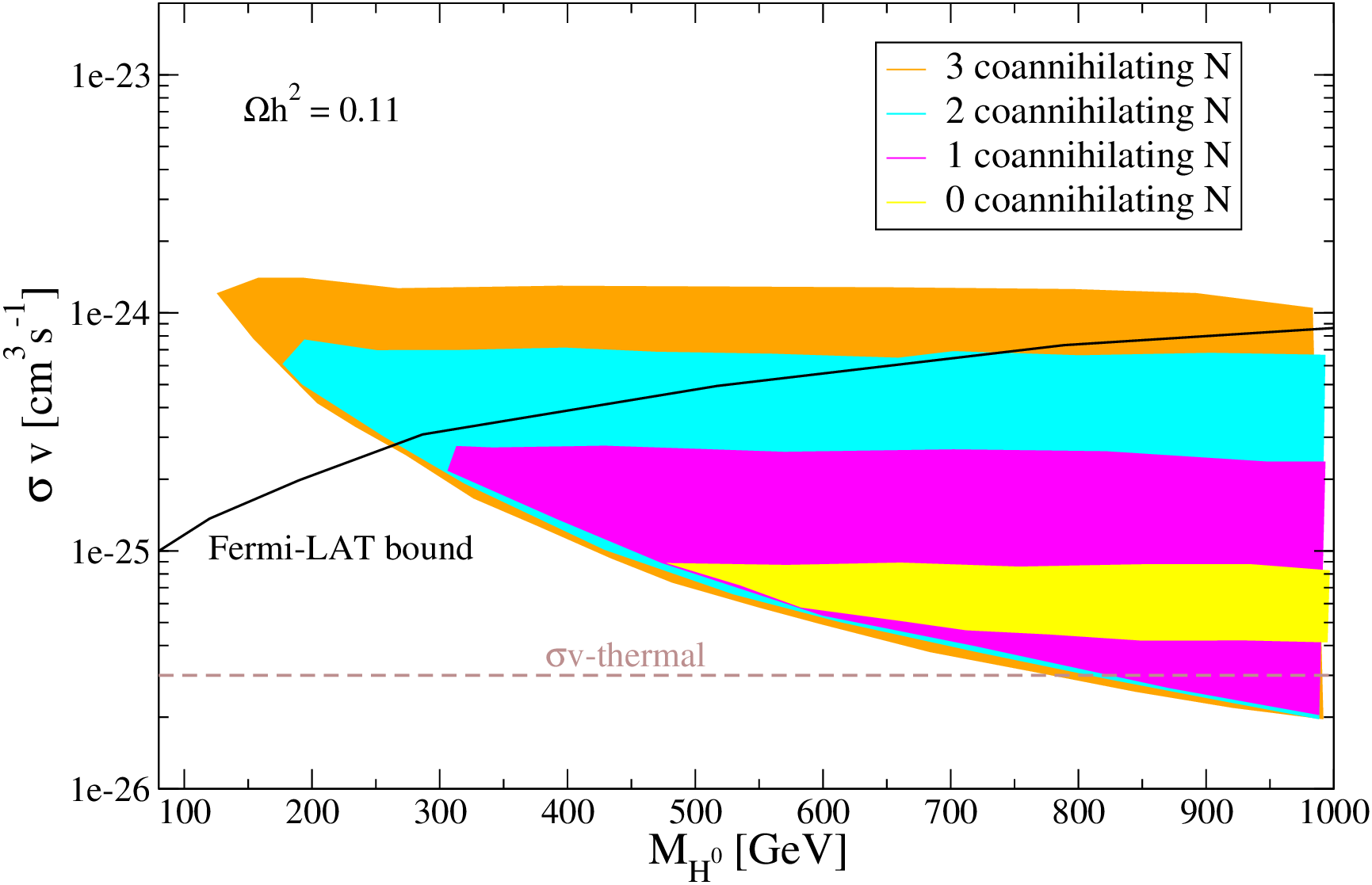} 
\caption{Regions in the plance ($\mh$,$\sv$) that are consistent with the dark matter constraint for different numbers of coannihilating $N$. \label{fig:sigmavdegB}}
\end{center}
\end{figure}
A comparison of the viable regions in the plane ($\mh$,$\sv$) is presented in figure \ref{fig:sigmavdegB}. The yellow region (the most internal one) shows the viable region in the absence of coannihilations with the right-handed neutrinos. That is, it is the result for the inert doublet model. Notice that in that case the minimum viable mass is  about $\mh\sim 500~\gev$ and that $\sv$ lies slightly above its thermal value, approximately between $4\times 10^{-26}\mathrm{cm^3s^{-1}}$ and $8\times 10^{-26}\mathrm{cm^3s^{-1}}$. As we include coannihilating neutrinos, the minimum value of $\mh$ decreases and the range of variation of $\sv$ increases. For $1$ coannihilating neutrino, the viable region starts at $\mh\sim 300~\gev$ and the maximum value of $\sv$ reaches  almost $3\times 10^{-25}\mathrm{cm^3s^{-1}}$. If all three neutrinos coannihilate with $H^0$, one can obtain viable models already at $\mh\sim 100~\gev$ and $\sv$ might be above $10^{-24}\mathrm{cm^3s^{-1}}$. More generally, this figure demonstrates that there is no incompatibility between a large value of $\sv$ (say of order $\text{few}\times10^{-25}\mathrm{cm^3s^{-1}}$) and the idea of a thermal freeze-out. Both can be reconciled via coannihilations in the early Universe.

Figure \ref{fig:sigmavdegB} also shows the current bound on $\sv$ from Fermi-LAT \cite{Ackermann:2012rg}. It excludes the low mass region, $\mh\lesssim 250~\gev$, and also the models with the largest values of $\sv$ over the entire mass range we consider. There are still, however,  sizable regions where the coannihilation effects are important that are not constrained by present data.  Future indirect detection data will certainly probe this region further, either improving the constraints or finding evidence of its existence. 

Summarizing, coannihilations with right-handed neutrinos may play a very important role in the dark  matter phenomenology of the radiative seesaw model. They open up new viable regions, change the expected mass spectrum, and modify the prospects for the direct and the indirect detection of dark matter.    

\section{Discussion}
\label{sec:dis}
We have throughout this paper assumed a specific structure for the matrix of neutrino Yukawa couplings, equation (\ref{eq:hs}). One may wonder, therefore, how our results depend on that particular choice. From the discussion on coannihilations, section \ref{sec:coan}, we know that for $N_i$ coannihilations to increase the relic density it must happen that $\sigma_{H^0H^0}\gg \sigma_{H^0N_i}$, $\sigma_{N_iN_i}$. So, all we need to reproduce our results is to ensure that $h_{\alpha i}$ are such that the above conditions on the cross sections are satisfied, and it  turns out that they usually are. In fact, we have repeated the scan over the parameter space of the model (see section \ref{sec:imp}) but allowing  the $9$ parameters $h_{\alpha i}$ to vary randomly without being subject to any constraint related to neutrino masses. The  results so obtained are indistinguishable from those presented here. We can thus conclude that  the effects discussed in this paper do not depend on the choice made in equation (\ref{eq:hs}) and, more generally, that they are not that sensitive to the experimental data on neutrino masses.

The discovery potential of the inert doublet at the LHC have been explored in the dilepton~\cite{Cao:2007rm,Dolle:2009ft},  trilepton~\cite{Miao:2010rg}, and multilepton~\cite{Gustafsson:2012aj} channels plus  missing energy. All these studies have focused on the light  mass regime $M_{H^0}\lesssim 80\ $GeV~\cite{Dolle:2009fn}. The high mass regime,  $M_{H^0}\gtrsim 500\ $GeV,  features small mass splittings between the odd scalars\cite{LopezHonorez:2006gr,Hambye:2009pw} and is out of the  LHC reach because of the small cross sections and rather soft leptons and jets. From figures~\ref{fig:mh0vsmasses}~and~\ref{fig:dmvsdmn}, we can see that now we can have substantially lower masses and much larger splittings. However, due to their large backgrounds,  we still expect the signals for $M_{H^0}\gtrsim 200\ $GeV to be difficult to  detect. The main effect of the coannihilating neutrinos at colliders in this scenario with scalar dark matter could be  the modification of the cascades leading to $H^0$. If some of the Yukawa couplings $h_{\alpha i}$ are competitive with the gauge couplings, and the spectrum is such that $M_{H^\pm}>M_{N_i}$  ($M_{A^0}>M_{N_i}$), then the leptonic signals (missing energy) could be increased through processes like $H^\pm\to l^\pm \nu H^0$ ($A^0\to \nu \nu H^0$) mediated by $N_i$. A more detailed study of these signals and of their possible relevance for LHC searches is left for future work.

As we have seen, in the radiative seesaw model, $N_i$ coannihilations allow to boost $\sv$ well above the thermal value of $3\times10^{-26}\mathrm{cm^3s^{-1}}$, enhancing in a significant way the indirect detection signals in this model. The idea of using coannihilations to increase the indirect detection rate is certainly not new; it was examined, for example, in \cite{Profumo:2006bx}. But it seems to us that, in spite of its possible relevance to current indirect detection data,  it has not received much attention lately. For that reason, we would like to emphasize that this setup is quite versatile and can be applied within different models of dark matter. The necessary ingredients are just two: a dark matter particle with a large self-annihilation cross section and  a number of coannihilating partners with weaker interactions.  The required number of coannihilating particles and their mass splitting is certainly model dependent but it can easily be calculated on a case by case basis. Notice that, among the different possibilities that have been considered to increase the dark matter indirect detection rate (gravitational clustering of dark matter, non-thermal production of dark matter, non-standard freeze-out, etc.), coannihilations seems to be the simplest one, allowing to preserve the idea of a thermal freeze-out and making definite predictions about  the particle spectrum of physics beyond the Standard Model.

\section{Conclusions}
\label{sec:con}
The radiative seesaw model is a simple extension of the SM that can account for neutrino masses and explain the dark matter. Within this model, we have considered the case in which the dark matter candidate is a scalar, $H^0$, and have studied the possible effect on the dark matter density of coannihilations with the right-handed neutrinos $N_i$. The dependence of these coannihilation effects with the relevant parameters of the model --the mass splitting, the  neutrino Yukawa couplings, and the number of coannihilating neutrinos-- was analyzed in detail. The net effect of the coannihilations with the right-handed neutrinos has been found to be an   \emph{increase} in  the relic density,  allowing to satisfy the dark matter constraint for smaller values of $\mh$. Its lower bound, in fact, moves from about $500~\gev$ (without $N_i$ coannihilations) to about $100~\gev$ (with all three $N_i$ coannihilating simultaneously).  Consequently, the dark matter phenomenology of this model can be substantially modified and may differ in important ways  from that of the inert doublet model. Making use of a random scan over the entire parameter space of this model, we have  investigated the new viable regions  that are obtained when coannihilations are taken into account. Such regions were shown to feature bigger values for the mass splitting between the scalar particles  and improved prospects for the direct and the indirect detection of dark matter.  Remarkably, one can obtain values of $\sv$ as large as   $10^{-24}\mathrm{cm}^3\mathrm{s}^{-1}$ (well above that normally associated with a thermal relic), with important implications for the indirect detection of dark matter.  Finally, we emphasized that one could use coannihilations  to reconcile large values of $\sv$ with a thermal freeze-out also within other models of dark matter.  

\section*{Acknowledgments}
The work of M.K.  and C.Y. is partially supported by the ``Helmholtz Alliance for Astroparticle Phyics HAP''
 funded by the Initiative and Networking Fund of the Helmholtz Association. DR and OZ have been partially supported by Sostenibilidad-UdeA and COLCIENCIAS through the grant number 111-556-934918. 

\bibliographystyle{hunsrt}
\bibliography{darkmatter_new}
\end{document}